\documentclass[aps,preprint,showpacs]{revtex4}

\usepackage{color}
\usepackage{graphicx}
\usepackage{dcolumn}
\usepackage{amssymb}

\begin{document}
%


\title{Theory of charge density wave depinning by electromechanical effect}
\author{P. Qu\'emerais}
\email{pascal.quemerais@neel.cnrs.fr}
\address{ Institut N\'eel, CNRS and Grenoble-Alpes University, BP 166, 38042 Grenoble Cedex 9, France}

\begin{abstract}
We discuss the first theory for the depinning of low dimensional, incommensurate, charge density waves (CDWs) in the strong electron-phonon (e-p) regime. Arguing that most real CDWs systems invariably develop a gigantic dielectric constant (GDC) at very low frequencies, we propose an electromechanical mechanism which is based on a local field effect. At zero electric field and large enough e-p coupling the structures are naturally pinned by the lattice due to its discreteness, and develop modulation functions which are characterized by discontinuities. When the electric field is turned on, we show that it exists a finite threshold value for the electric field above which the discontinuities of the modulation functions vanish due to CDW deformation. The CDW is then free to move. The signature of this pinning/depinning transition as a function of the increasing electric field can be directly observed in the phonon spectrum by using inelastic neutrons or X-rays experiments.
\end{abstract}


\maketitle

\section{Introduction }

Charge density waves (CDWs) are broken symmetry states of low-dimensional metals, induced by the e-p interaction (for a recent review, see \cite{review1D_4}). The CDW ground-states generally compete with other electronic instabilities, such as superconductivity, spin density waves, or excitonic insulators. For that reason, it remains essential to understand their origin and their physical properties. Many experiments show evidences of strong e-p coupling in a lot of one and two-dimensional CDW systems \cite{monceau_strong, weber_2, maschek, theory_Rte}. In this respect, the new quasi two-dimensional rare-earth tritellurides (ReTe$_3$, where Re$=$rare-earth) appear nowadays among the most convaincing. For this family of compounds and in the low temperature CDW state, the ratio of the electronic gap $2 \Delta$ to the critical temperature $T_c$ is about $15$ \cite{moore,sacchetti}, which is more than 4 times the predicted weak coupling value which is $3.5$. The modulation function of the CDW has also been experimentally studied in details for these compounds \cite{malliakias_1,malliakias_2,malliakias_3}. It was shown that the function has nothing to do with the weak coupling prediction, i.e. a simple cosine-like modulation. On the contrary, discontinuities appear in this function, indicating the formation of local bonds, oligomers and discommensurations as expected in the strong coupling theory \cite{aubry_ledaeron,aubry_quemerais,aubry_raimbault}. The Kohn anomaly  \cite{review1D_3} in the phonon spectrum also do not behave as expected by the weak coupling theory. The spectrum was measured by inelastic X-rays \cite{maschek}. A gap of about $5$meV persists in the low temperature phase although the compound is incommensurate. Moreover, the softening of the phonons is large and spreads out over a large part of the Brillouin zone indicating a strong e-p interaction \cite{maschek, theory_Rte}.

At low temperatures, the CDWs in the quasi-one dimensional compounds NbSe$_3$, TaS$_3$, A$_{0.3}$MoO$_3$ ( A=K,Rb), (TaSe$_4$)$_2$I, carry an extra non-linear current above a certain threshold electric field $E>E_c$ of the order of $0.1-1$V/cm \cite{review1D_1, review1D_1_2, review1D_1bis, review1D_1_3, gorkov_gruner, thorne_review, review1D_4}. Recently, the same phenomenon has been observed in the two dimensional ReTe$_3$ family \cite{sinchenko_1,sinchenko_2,sinchenko_3,lebolloch_1}. All the available theories of the electrical depinning of incommensurate CDWs are build within the weak e-p coupling scheme. In this framework, the residual impurities are the unique source of pinning, commensurability effects being negligible in these compounds. By contrast, at strong coupling, it is the lattice itself which is the main source of the CDW pinning. Aubry et al. \cite{aubry_ledaeron,aubry_quemerais,aubry_raimbault} showed that there is a  pinning/depinning transition as a function of the strength of the e-p interaction in incommensurate Peierls chains. This pinning mechanism is intimately related to the presence of discontinuities in the modulation function of the atomic positions. At weak coupling, the continuity of this function assures the existence of the Fr\"ohlich mode since the energy of the condensate remains independent of the CDW phase. Owing to this Fr\"ohlich mode, the CDW state remains metallic unless impurities pin the CDW phase. At strong coupling, discontinuities appear in the modulation function of the ground-state and the Fr\"ohlich mode is lost, leading to a dielectric state, independently of the presence or not of impurities. Importantly, the energy scales of the two different pinning mechanisms, due to the impurities at weak coupling, or due to the lattice at strong coupling are very different. In the first case, the pinning energy at a bond length scale has been estimated to be of the order of $10^{-7}$ eV \cite{review1D_1_2}. By contrast, the lattice pinning yields energies of the order of a few hundred Kelvin (i.e. $10^{-3}$ eV) at the same length scale \cite{aubry_quemerais}. This makes a difference of several orders of magnitude in energy between both mechanisms.

So far, a depinning mechanism under an electric field in the strong coupling framework is missing. In this letter, we propose to fill in this theoretical gap. We restrain ourselves in the paper to one-dimensional (1D) Peierls chains. More precisely, we consider a sample made of a three dimensional assembly of linear Peierls chains, all oriented in the same direction. 

\section{Model for the Peierls chains}
To illustrate our aim, we choose the Sue-Schrieffer-Heeger (SSH) tight-binding hamiltonian \cite{SSH} to describe a single chain,

\begin{eqnarray} \label{SSH_1}
 H_{ssh} &=& -\sum_{n, \sigma} \left( t_0 -\alpha (u_{n+1}-u_n) \right)
(a_{n+1, \sigma}^\dag a_{n,\sigma} + h.c.) \nonumber \\ 
&+& \frac{K}{2} \sum_n (u_{n+1}-u_n-b)^2.
\end{eqnarray}

The first term is the electron hamiltonian, where $a_{n,\sigma}$(respectively $a^\dag_{n,\sigma}$) is the annihilation (respectively creation) operator of an electron with spin $\sigma$ at the $n^{th}$ site along the chain. $t_0$ is the bare electronic hopping integral. $\alpha$ is the electron-phonon coupling. The $\{u_n \}$ give the atomic coordinates along the chains, and we introduce $v_n=u_{n+1}-u_n$, the bond length between the atom $n$ and $n+1$. The electron hopping integral depends on these bond lengths. Finally, the second term in (\ref{SSH_1}) is the elastic energy of the chains, $K$ being the elastic constant.

When $\alpha=0$, the energy is minimal when the atoms are equidistant with $u_n-u_{n-1}=b$. In the following, we consider a chain of $N$ atoms with $P$ pairs of electrons (i.e. $P$ electrons with spin up and $P$ electrons with spin down). If $\alpha \ne 0$, we fix to $a$ the interatomic distance $v_n=u_n-u_{n-1}=a$ in the undistorded state. The parameter $b$ which minimizes the energy becomes $b=a+ (4 \alpha/K \pi) \sin(\pi P/N)$.

It is convenient to introduce the following dimensionless parameters and variables: $t=t_0-\alpha a$, $\lambda=\alpha \sqrt{2/Kt}$, $V_n=v_n \sqrt{K/2t}$ and $B=(2 \lambda/\pi) \sin{(\pi P/N)}$. The variable $V_n$ is the dimensionless length of the bond between the $n^{th}$ and ${n+1}^{th}$ atom. In the bond order wave (BOW) state, which is the type of CDW state developed in the SSH model, the bond lengths $\{V_n\}$ are modulated. 

\begin{figure}
\includegraphics[scale=0.50]{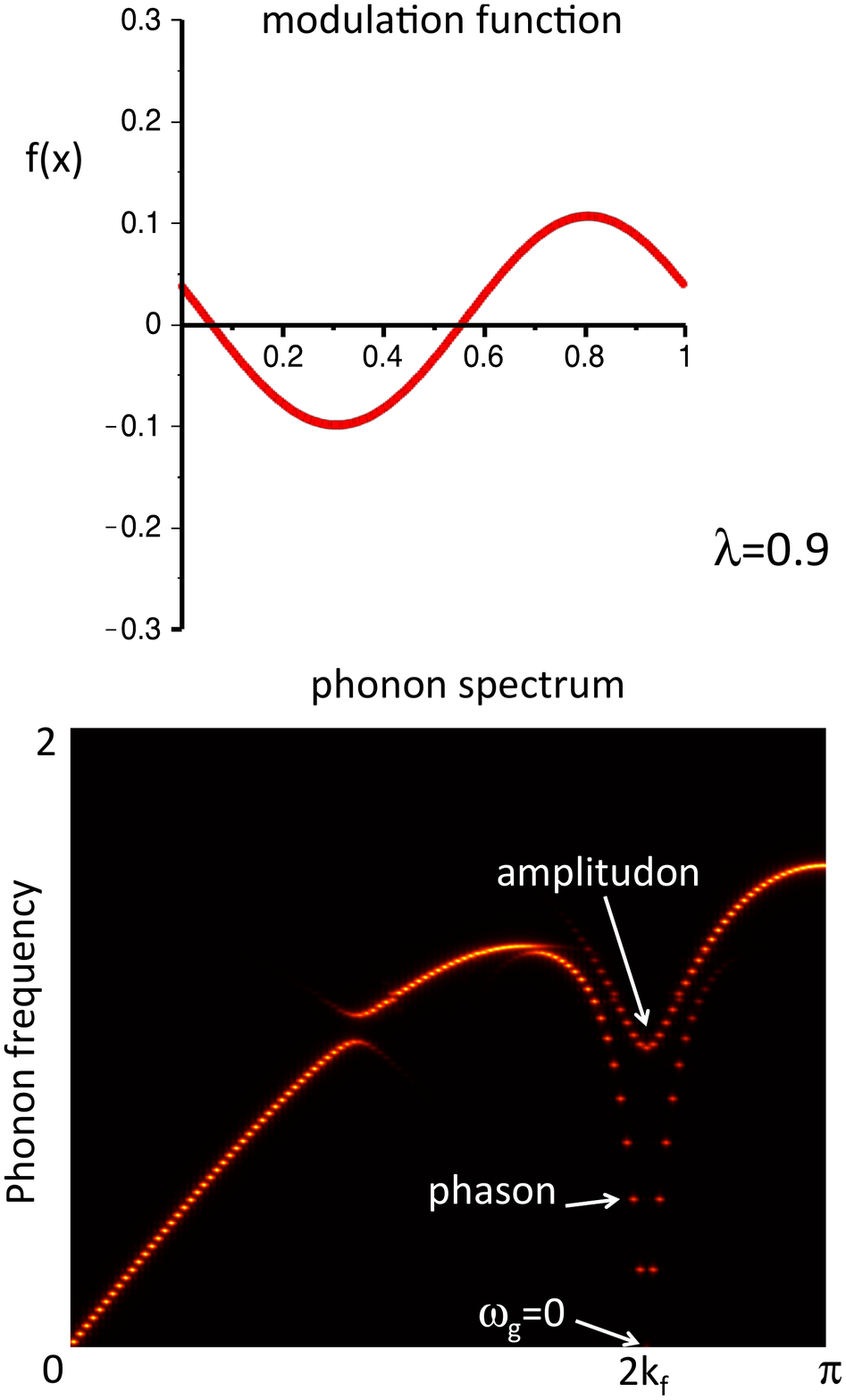}
\caption{Modulation function and mapping of the phonon spectrum in the weak e-p coupling regime. $\lambda=0.9$ and $\zeta_{\ell}=89/233$. The modulation function $f(x)$ is continuous, indicating the existence of the Fr\"ohlich mode. The calculated phonon spectrum is characterized by a zero phonon frequency ($\omega(2k_f)=\omega_g=0$). The phase and amplitude modes around $2k_f$ are visible on the mapping.}
\label{fig.1}
\end{figure}

\begin{figure}
\includegraphics[scale=0.50]{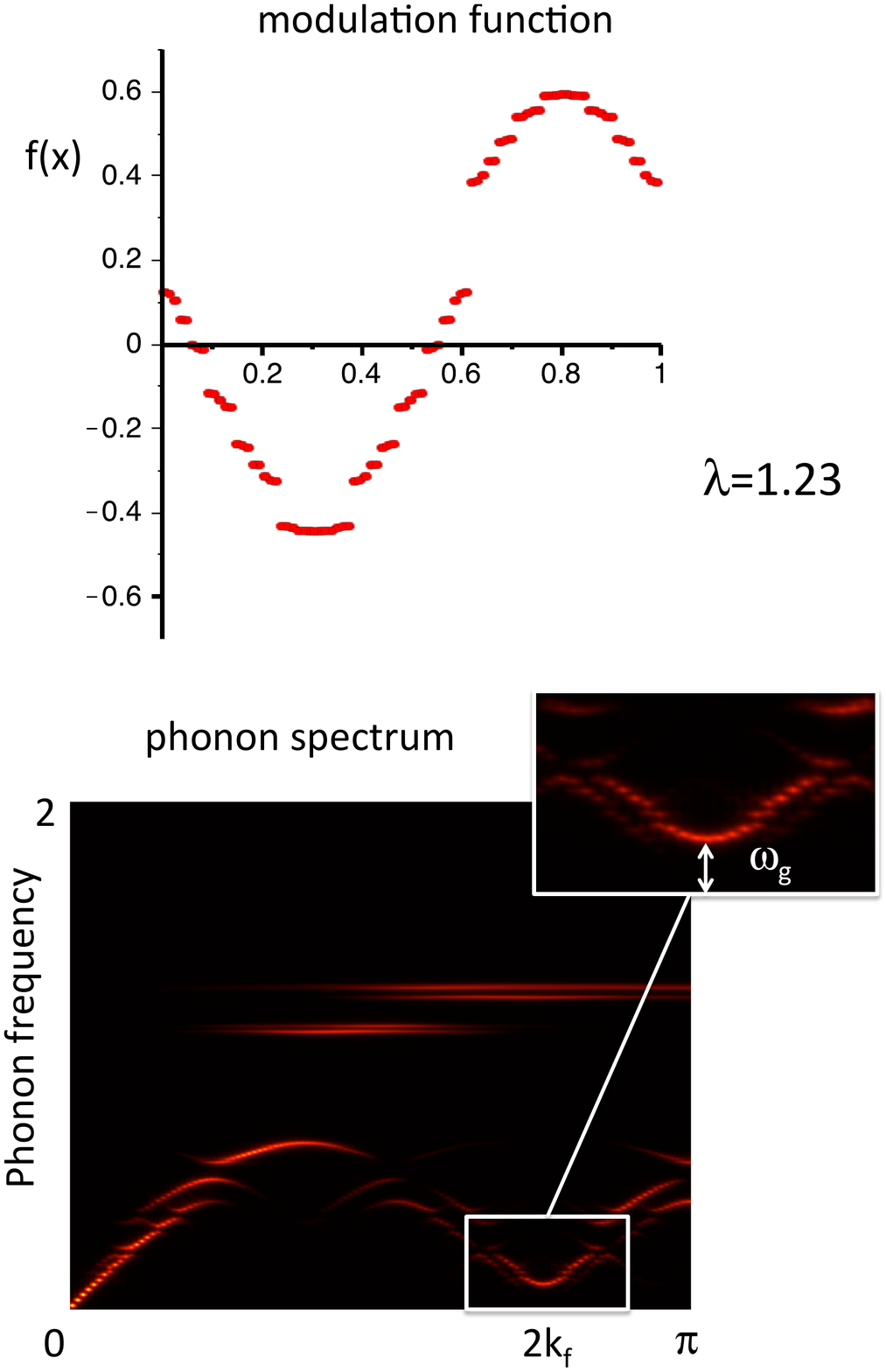}
\caption{Same figure as Fig.1 in the strong e-p regime, $\lambda=1.23$. The modulation function is discontinuous (see text). The phonon softening concerns a large part of the Brillouin zone around $2k_f$, and there is a pinning gap $\omega_{2k_f}=\omega_g \ne 0$.}
\label{fig.2}
\end{figure}

The total energy $\Phi$ of the chain which is the sum of the electronic and the elastic energy may be written by using these new parameters and variables,

\begin{eqnarray} \label{energy}
\Phi \left( \{V_n\} \right)/2t = \sum_\mu E_{\mu}+ \frac{1}{2} \sum_n (V_n-B)^2.
\end{eqnarray}

The sum over $\mu$ runs for the $P$ lowest occupied states (at $T=0$K), and the corresponding energies $E_{\mu}$ satisfy the eigenequations ($H_{ssh} \mathbf{\Psi}_{\mu}=E_{\mu} \mathbf{\Psi}_{\mu}$),

\begin{equation} \label{eigen}
-(1-\lambda V_{n+1}) \Psi_{n+1}^{\mu}-(1-\lambda V_{n-1}) \Psi_{n-1}^{\mu}=E_{\mu} \Psi_n^{\mu},
\end{equation}

where $\mathbf{\Psi}^{\mu}=^t(\Psi_1^{\mu} \cdots \Psi_N^{\mu})$ is the electronic wave function in the state $\mu$. The atomic positions in the ground-state are numerically calculated by minimizing the total energy with respect to the bond lengths.

\begin{equation}
\label{gradient}
\frac{\partial \Phi}{\partial V_n} = V_n-B+2 \lambda \sum_{\mu=1}^{P} \Psi_n^{\mu} \Psi_{n+1}^{\mu}.
\end{equation}
 
The bond lengths in the ground-state are found by using (\ref{gradient}) and a usual gradient procedure. The calculations are performed with a precision of $10^{-6}$. Finally, the bond lengths can be represented by the modulation function $f(x)$ defined by $V_n=f(x_n)$ for all $n$, where $x_n=(nP/N+ \phi)$ modulo $1$, and $ 2 \pi \phi$, with $0 \le \phi \le 1$, is an arbitrary phase. The modulation function is periodic, $f(x+1)=f(x)$, $0\le x \le 1$. For two numbers $P$ and $N$ prime one to another, the function $f(x)$ in the ground-state is thus defined with $N$ equidistant points on the segment $[0,1]$.

The parameter $\zeta=P/N$ is the commensurability ratio of the considered Peierls chain and is related to the Fermi wave vector $k_f=\pi \zeta$. For an incommensurate system, $\zeta$ must be an irrational number. As an example, we will choose $\zeta=\zeta_{\infty}=(3-\sqrt{5})/2 \approx 0.381966 \cdots$. To simulate such a system, the usual procedure is to approximate $\zeta_{\infty}$ by a succession of rational numbers $\zeta_1$, $\zeta_2 \cdots$, $\zeta_{\ell}$, $\cdots$, with $\lim_{\ell \rightarrow \infty} \zeta_{\ell}=\zeta_{\infty}$.These numbers are obtained by the theory of Diophantine approximations of irrational numbers \cite{diophantine}. As successive approximants of $\zeta_{\infty}$, we get the commensurability ratios $P_{\ell}/N_{\ell}=$, $1/2$, $2/5$, $3/8$, $5/13$, $8/21$, $13/34$, $21/55$, $34/89$, $55/144$, $89/233 \cdots$. In practice, commensurability effects exponentially decreases with $N_{\ell}$. Consequently, commensurate systems with commensurability order $N_{\ell} \sim 50-100$ (with $P_{\ell}$ and $N_{\ell}$ prime one to another) are largely sufficient to study incommensurate systems.

Besides the modulation function, the calculation of the phonon spectrum in the CDW state is of prime importance since it can be experimentally measured by inelastic neutrons and/or inelastic X-rays studies. The eigenvalues and the eigenstates $\omega_{\nu}$ and eigenvectors $\mathbf{e}_{\nu}=(e_n^{\nu})$ for  the phonons are obtained by diagonalization of the dynamical matrix $\mathbf{M}= (M_{m,n})_{1 \le m,n \le N}$, where $M_{m,n}= \partial^2 \Phi/\partial u_n \partial u_m$. We first calculate \cite{ledaeron_thesis,aubry_quemerais},

\begin{equation}
A_{m,n}= \frac{\partial^2 \Phi}{\partial V_n \partial V_m}= \delta_{m,n}-2\lambda^2 \sum_{\mu_{o.},\nu_{u.}} \frac{\Delta_{m,n}^{\alpha,\beta}}{E_{\beta}-E_{\alpha}}.
\end{equation}
The sum over $\nu_{u.}$ (respectively $\mu_{o.}$) runs for the unoccupied electronic states (resp. occupied states). Moreover,
\begin{equation}
\Delta_{m,n}^{\alpha,\beta}= \left(\Psi_{n+1}^{\beta}\Psi_n^{\alpha}+\Psi_{n+1}^{\alpha}\Psi_n^{\beta}\right) \left(\Psi_{m+1}^{\beta}\Psi_m^{\alpha}+\Psi_{m+1}^{\alpha}\Psi_m^{\beta}\right)
\end{equation}
$M_{m,n}$ is then obtained from 
\begin{equation}
M_{m,n}=\left[A_{m,n}+A_{m+1,n+1}-A_{m-1,n}-A_{m,n-1}\right].
\end{equation}

The one-phonon dynamical scattering function $S(q,\omega)$ is calculated from the knowledge of $(\omega_{\nu}, \mathbf{e}_{\nu})$ \cite{ledaeron_thesis,aubry_quemerais}.
\begin{equation}
\label{spectrum}
S(q,\omega) = \sum_{\nu}{\left[\sum_{i,j} e_i^{\nu}e_j^{\nu} \cos{(q(u_i-u_j)} \right] p(\omega,\omega_{\nu})}, 
\end{equation}
where

\begin{equation}
p(\omega,\omega_{\nu})= \frac{n(\omega)}{\omega_{\nu}} \delta(\omega-\omega_{\nu})+\frac{n(\omega)+1}{\omega_{\nu}} \delta(\omega+\omega_{\nu}).
\end{equation}

$n(\omega)$ is the Bose factor $[e^{-\hbar \beta \omega}-1]^{-1}$. However the sum in (\ref{spectrum}), whose terms are proportional to $n(\omega)/\omega_{\nu}$, is not convenient to represent the whole phonon spectrum since it strongly favorizes the low energy phonons. To get a better representation of the whole spectrum, we introduce the damped harmonic oscillator function,
\begin{equation} \label{poids}
p(\omega,\omega_{\nu})= \frac{\Gamma^2}{(\omega-\omega_{\nu})^2+ \Gamma^2}.
\end{equation}

$\Gamma$ is a small empirical parameter ($\sim 0.005$) which smoothes the discrete sum over $\nu$. 
The function $S(q,\omega)$ calculated by using (\ref{poids}) behaves similarly to the true $S(q,\omega)$ (the phonons are at the same location in the plane $(q,\omega)$) but gives an homogeneous, temperature independent, mapping of the whole spectrum.

It has been shown by Le Daeron and Aubry \cite{aubry_ledaeron} that for a given irrational $\zeta$, it exists a finite e-p coupling $\lambda_c(\zeta)$, which separates the weak from the strong coupling regime. That transition, originally called the transition by breaking of analyticity (TBA), occurs for $\zeta_{\infty}$ at $\lambda_c \approx 1.21$, coupling at which discontinuities appears in the modulation function.
The Figs.1-2 show the modulation function and the related phonon spectrum for $\zeta = 89/233 \approx \zeta_{\infty}$ for two e-p coupling values. At weak coupling Fig.\ref{fig.1}, the modulation function is continuous. Owing to this continuity, the condensate energy is independent of the CDW phase $\phi$. The CDW is not pinned and is free to move. In this regime, the Fr\"ohlich collective mode ($d\phi/dt \ne 0$) allows the transport of electrons. The related phonon spectrum develops a complete Kohn anomaly with $\omega(2k_f)=0$, and the softening of the phonons is restrained around $2k_f$. The original phonon branch is replaced around $2k_f$ by a separated phason (CDW phase excitation) and amplitudon (CDW amplitude excitation) which may be observed on the spectrum.
By contrast, at strong coupling (Fig.\ref{fig.2}), the modulation function presents discontinuities. This is the signature of the lock in of the CDW phase, which is pinned by the lattice, although the system is incommensurate. The CDW is not free to move and the Fr\"ohlich mode is lacking. The related phonon spectrum also shows a behavior different than in the weak coupling regime. First, there is an energy gap at $2k_f$ in the spectrum: $\omega(2k_f) \ne 0$. Second, the softening of the phonons concerns a large part of the Brillouin zone around $2k_f$. It is interesting to note that these two properties : existence of a phonon gap at $2k_f$ and a large spreading of the phonon softening over the Brillouin zone have been observed for (TaSe$_4$)$_2$I \cite{neutron_tase4_2i}, but also for the Re-Te$_3$ family \cite{maschek} and 2$H$-NbSe$_2$ \cite{weber_2}. For NbSe$_3$, no softening could be observed \cite{inelas_nbse3}, and for K$_{0.3}$MoO$_3$ a phonon gap persists below the transition temperature \cite{neutron_kmo3,xray_kmo3}. All these experimental observations are in favor of the strong coupling regime. More details about the transition by breaking of analyticity may be found in \cite{aubry_quemerais}.

\section{Electromechanical effect}

To understand how a CDW in the strong e-p regime may be depinned from the lattice by an electric field, we must take into account one of the most amazing experimental properties: the gigantic low frequency dielectric constants of CDW systems. In the microwave range of frequencies and below ($\omega \sim < 100$MHz), all known incommensurate one-dimensional CDW compounds behave as dielectrics at low temperatures with dielectric constants as large as $10^{6}-10^{9}$. We note that for the two-dimensional Re-Te$_3$ family, such measurements in the microwave range are still lacking. In what follows, due to this dielectric behavior, we assume that the possible presence of free electrons (even at $T=0$K which is the case for NbSe$_3$ and Re-Te$_3$) is not sufficient to screen such a high polarization. We completely neglect their effect and treat the CDW states as pure dielectrics.

Since in the strong e-p coupling regime, the CDW is pinned by the lattice, each bond of the Peierls chains polarizes under the action of an applied electric field $\mathcal{E}$. The induced dipolar momentum of the bond between $n$ and $n+1$ atom may be written as 
\begin{equation} \label{dipole}
p_n= \alpha_n \mathcal{E},
\end{equation}
where $\alpha_n$ is the polarizability of the $n^{th}$ length. We don't know the exact value of this quantity, but we can derive some general properties. First, it is important to emphasize that the typical polarizability of a bond is always very small $\sim 10^{-38}-10^{-40}$Cm$^{2}$/V \cite{israel}. Second, the polarizability of a given bond varies with its length, a property which is used in the semi-classical theory of the Raman effect \cite{raman}. Third, in the ground-state (i.e. $\mathcal{E}=0$), the bonds, whose lengths are given by the set $\{v_n^{gs}\}$ do not carry any dipolar momentum. Finally, one may expand the polarizability as a function of the bond length as

\begin{equation} \label{polarizability}
\alpha_n \approx \alpha^{(0)}+ \alpha^{(1)} (v_n-v_n^{gs}).
\end{equation}

$\alpha^{(0)}$ is a term which accounts for the ionic polarizability. It will not play any role in the following. $\alpha^{(1)}$ is the variation of polarizability with the bond length variation. It may be estimated to about a few tens of $10^{-40}$Cm$^{2}$/V per $\AA$ \cite{raman}, i.e. $10^{-28}-10^{-30}$Cm/V. Its sign is of importance. Since in the CDW state the electronic density participating to a given bond becomes larger when its length decreases, we assume that $\alpha^{(1)}<0$: a shorter bond gives a larger polarizability. We also need the density of bonds in the 3D compounds. On a single chain, the mean distance $\bar v$ between two consecutive atoms is of the order of a few $\AA$. Let us fix $\bar v=3 \AA$. For simplicity, we also assume that the atoms in the undistorded state (above $T_c$) and in the two other directions are arranged in a cubic symmetry. The density of bonds oriented in the chains direction is thus $n \approx 10^{29}m^{-3}$.

In usual experimental settings for electrical measurements, a voltage is applied to the CDW sample of length $L$. In that case, the \textit{macroscopic} field $E=V/L$ inside the sample is \textit{imposed}. When such a macroscopic electric field is applied in the direction of the chains, \textit{all bonds along the chains in the sample polarize in the same direction}. For a given bond $i$ of a chosen (central) chain, the dipolar energy $E_{dip}(i)$ involved, which includes the dipole-dipole interactions, may be written as

\begin{equation}
E_{dip}(i) = -p_i \left[E- \frac{1}{4 \pi \epsilon_0}\left( \sum_{j \ne i} \frac{3 (\mathbf{e}_x \cdot \mathbf{e}_{i,j})-1}{\vert \mathbf{r}_{i,j} \vert ^3} \right)p_j \right].
\end{equation}

The index $j$ runs for all the other bonds in the 3D sample. $\epsilon_0$ is the permittivity of the vacuum. $\mathbf{r}_{i,j}=\mathbf{r}_i-\mathbf{r}_j$, where $\mathbf{r}_i$ and $\mathbf{r}_j$ are the cooordinates of the bonds $i$ and $j$ respectively. $\mathbf{e}_x$ is the unit vector in the chain direction, and $\mathbf{e}_{i,j}=\mathbf{r}_{i,j}/ \vert \mathbf{r}_{i,j} \vert$. In the mean field approximation, we may replace $p_j$ by its mean values $<p_j>=\bar p$ for all $j \ne i$. That leads to the well-known Lorentz local field results \cite{kittel}, which, in a cubic symmetry, leads to

\begin{equation}
\frac{\bar p}{4 \pi \epsilon_0}\left( \sum_{j \ne i} \frac{3 (\mathbf{e}_x \cdot \mathbf{e}_{i,j})-1}{\vert \mathbf{r}_{i,j} \vert ^3} \right) =P/3 \epsilon_0,
\end{equation}
where $P$ is the sample polarization, which is related to the macroscopic field $E$ in the sample by $P=\epsilon_0(\epsilon-1) E$. Finally,

\begin{equation}\label{energy_dipole}
E_{dip}(i) = -p_i \left( \frac{\epsilon+2}{3} \right) E.
\end{equation}
 
Owing to the very large dielectric constant $\epsilon$ of a CDW compound, \textit{the local field is orders of magnitude larger than the macroscopic field $E$ in the sample}. It is of the order of $10^8-10^{10}$V/m. For comparison, an electron at about 1$\AA$ creates a  \textit{microscopic} field $\vert \mathbf{e} \vert\sim10^{11}$V/m.

In the following we argue that the forces created by this local field behave as pressure forces which are sufficient: i) to induce a significant deformation of the CDW; ii) to close the discontinuities of the ground-state modulation function; and finally iii) to depin the CDW.

Let us estimate the dipolar momentum (\ref{dipole}) and the energy (\ref{energy_dipole}) involved under the local field. Since $P=n \bar{p}$, we obtain $\bar p \sim 10^{-30}$Cm ($0.3$ Debye), by taking $E=1$V/cm, $\epsilon=10^8$ and $n=10^{29}$m$^{-3}$. The resulting energy (\ref{energy_dipole}) is $E_{dip} \sim 0.06$eV. This is roughly a few percents of the amplitude of the other terms involved in the original hamiltonian (\ref{SSH_1}). It is sufficient to compete with them and to induce a small variation $\Delta v_n$ of the bond lengths. We may also estimate the order of magnitude of this variation. From (\ref{dipole}) and (\ref{polarizability}), $\bar{p} \sim \alpha^{(1)} \Delta v E_{loc}$. With $ \alpha^{(1} \sim -10^{-29}$Cm/V, we get a mean contraction of the bond lengths of $\Delta \bar v \sim -10^{-2} \AA$. This is a piezoelectric effect. In reality, it can be observed only if one of the ends of the sample is free to move, otherwise the contraction of a part of a chain leads necessary to the stretching of the remaining part. That phenomenon may induce a supplementary long wavelength deformation of the CDW which is not included in our model (the chains here are supposed free to contract). Is is interesting to note that such a piezoelectric effect has been observed for TaS$_3$ with a very large piezoelectric constant $> 10^{-4}$ cm/V \cite{pokrovskii}. Within our crude estimation ($\Delta \bar v=10^{-2} \AA$, $\bar v=3 \AA$ and $E=1$V/cm), we get $\sim 10^{-3}$cm/V as piezoelectric constant. 

The electromechanical effect invoked in the title of the article takes its origin here: the alignment of the created dipoles under the electric field, which is imposed by the quasi-one unidimensional electronic structure, associated with a large enough density of induced dipoles, allows the system to have a gigantic dielectric constant. Although the amplitude of the dipoles is rather moderate, it is sufficient to create a huge local field. In turn, this local field acts as a pressure force on the chains which is sufficient to change the bond lengths and to induce a CDW deformation.

As discussed above, at the scale of a bond length, it is the local field $E_{loc} \sim (\epsilon+2)E/3$ which is felt by the polarizable bonds. Under the electric field $E$, the resulting hamiltonian for a unique chain then becomes,

\begin{equation}
H(E)=H_{ssh}-\sum_i \alpha^{(1)} \left(\frac{\epsilon+2}{3}\right)^2 E^2 \left( v_i-v_i^{gs}\right).
\end{equation}

In our system of units, the total energy of a single chain under an applied voltage becomes,

\begin{equation}
\label{energie_E}
\Phi(\{V_n\},E)/2t = \sum_\mu E_{\mu}+ \frac{1}{2} \sum_n (V_n-\tilde{B}(E))^2+C,
\end{equation}

\begin{figure}
\includegraphics[scale=0.5]{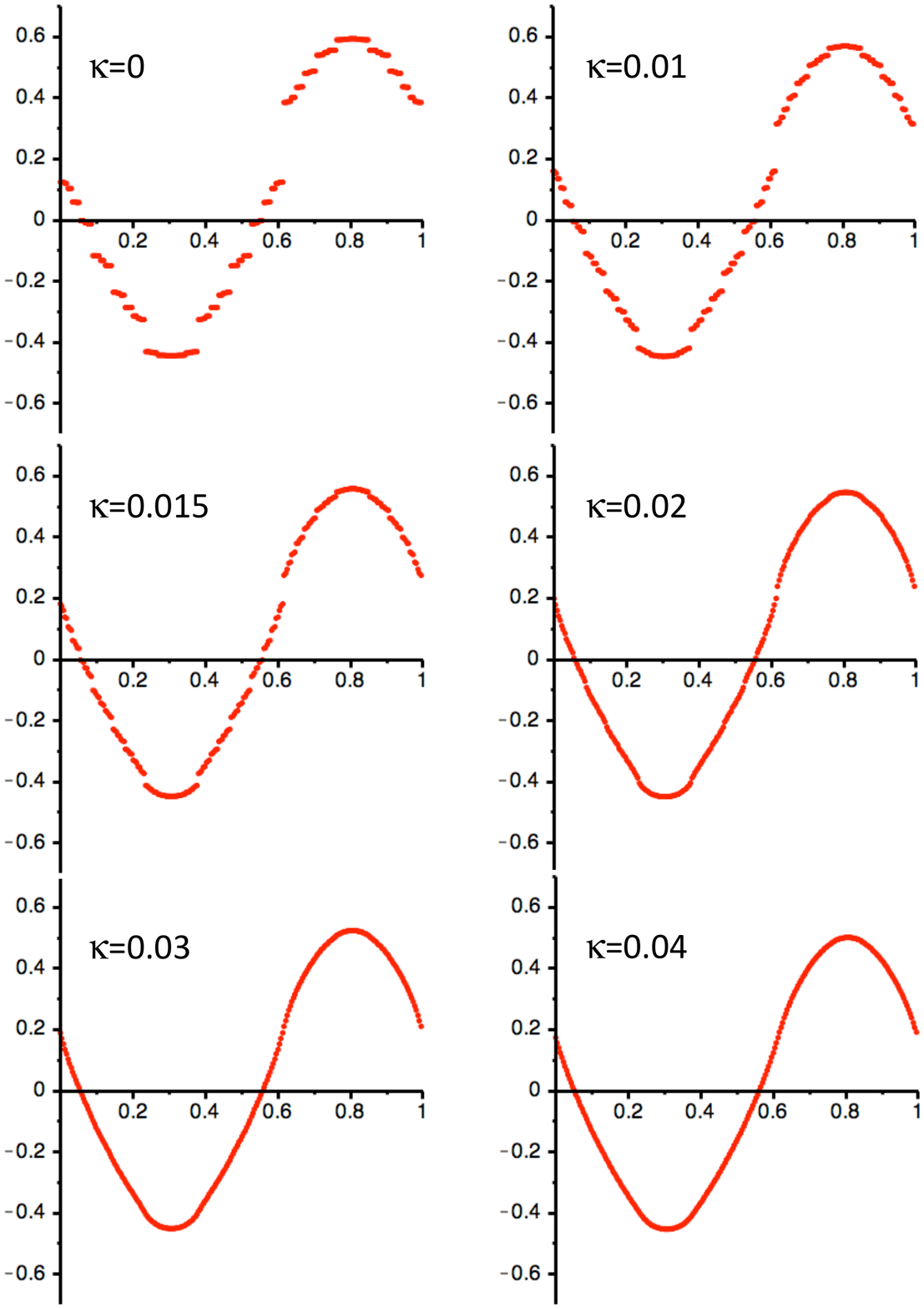}
\caption{Deformation of the CDW modulation function as a function of the electric field $\kappa$. The e-p coupling is fixed to $\lambda=1.23$. The initial discontinuities in the ground-state ($\kappa=0$) progressively disappear as the field increases. At about $\kappa=\kappa_c \approx 0.03$, the CDW is depinned from the lattice and free to move. The band filling is $\zeta_{\ell}=89/233$.}
\label{fig.3}
\end{figure}

with $\tilde{B}(E)=B+ \kappa(E)$, and

\begin{equation}
\kappa(E)= -\alpha^{(1)} \left(\frac{\epsilon+2}{3} \right)^2 E^2 \sqrt{\frac{1}{2tK}}.
\end{equation}

$C=C(E,\{v_i^{gs}\})$ is a constant term at fixed electric field $E$ which does not play any role in the CDW deformation.
The parameter $\kappa(E)$ may vary significantly, depending on the dielectric constant, the electric field and $\alpha^{(1)}$.

\begin{figure}
\includegraphics[scale=0.5]{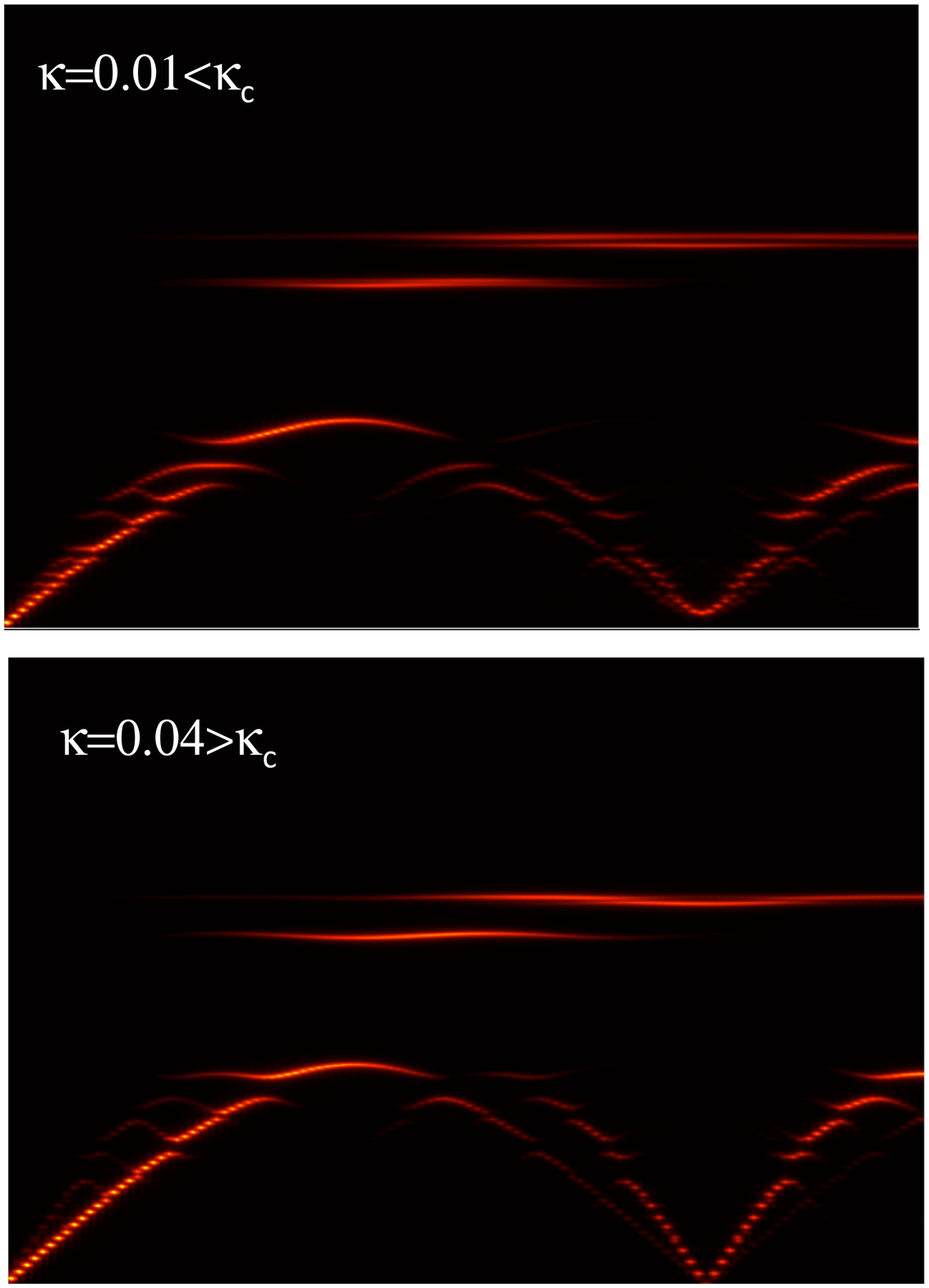}
\caption{Phonon spectra for two values of the electric field (see text). Below $\kappa_c$, the phonon gap decreases but persists. The CDW is still pinned by the lattice. Above $\kappa_c$, the pinning potential collapses leading to the closure of the phonon gap. The e-p coupling is $\lambda=1.23$ and the band filling $\zeta_{\ell}=89/233$.}
\label{fig.4}
\end{figure}

The minimization of (\ref{energie_E}) leads to the new bond lengths under the field $E$. Owing to the contraction induced by the electric field (through $\tilde B(E)$), the electronic hopping terms in (\ref{SSH_1}) increase, allowing the electrons to pass more easily from one site to another. This leads to the progressive closure of the discontinuities in the modulation function. The Fig.\ref{fig.4} shows its evolution as a function of the electric field $E$ ($\kappa(E)$). The depinning transition (when all the discontinuities of the modulation function vanish) occurs at about $\kappa_c \approx 0.03$. By taking the following reasonable parameters : $\epsilon=10^{8}$, $t=1$eV, $K=1$eV/$\AA^2$ and $\vert \alpha^{(1)} \vert=10^{-29}$Cm/V, the critical value $\kappa_c=0.03$ leads to an electric threshold field of $E_c=0.8$V/cm, which is a correct order of magnitude with respect to the experiments.

Eventually, the consequence of the present depinning mechanism is the important changes in the phonon spectrum. The Fig.\ref{fig.5} shows the phonon spectrum for two different non zero electric fields, below and above the threshold $\kappa_c$. Notice that the zero field reference spectrum is the one represented on the Fig.\ref{fig.2}. The spectra are largely modified around $2k_f$ by the application of the electric field. There is a global softening of the frequencies around $2k_f$. That phenomenon lowers the elastic constants of the chain, and thus the Young modulus. We note that such a progressive lowering of the Young modulus has also been observed \cite{brill}. Above the threshold $\kappa_c$, the Kohn anomaly is complete, the pinning potential collapses, the phonon gap vanishes and the Fr\"ohlich mode is restored. This is in contrast with the weak e-p coupling depinning scenario for which no important changes in the phonon spectrum are expected under electric field since the pinning potential due to the impurities remains the same whatever the applied electric field is.

\section{Discussion and conclusion}

In this paper, we have proposed the first depinning mechanism of CDWs in the strong e-p regime. It is based on the effect of the local field which is very large owing the huge dielectric constant of the compounds. Although it leads to a coherent CDW depinning scenario, many important points remain to be discussed with respect to the experiments. We mention here three of them.

First, it is experimentally observed that  for quasi-one dimensional systems, two electrical threshold fields (called $E_{c1}$ and $E_{c2}$) must be considered at low enough temperatures. In spite of numerous studies performed since more than three decades, their origin remains mysterious \cite{thorne1,thorne2}. $E_{c1}$ is the true onset of non-linear conductivity, and it is has been demonstrated that the induced current corresponds to a conduction by temporally ordered collective creeps which are thermally activated \cite{lemay}. The second threshold $E_{c2}$ corresponds to an abrupt depinning of the CDW leading to an increase of the non-linear current by several orders of magnitude. In our scenario, the threshold field which has been discussed corresponds to $E_{c2}$ at which the pinning potential collapses. In our view, $E_{c1}$ is related with the thermal excitations of the local (equidistant) discommensurations which naturally exist in the ground-state of the strong e-p coupling regime\cite{aubry_quemerais}.

Second, the SSH model that we have used to illustrate our scenario contains all necessary ingredients to describe qualitatively the physics involved. However it is quantitatively unsatisfactory as it leads to modulation amplitudes which are too large with respect to those observed in real systems \cite{pouget}. Another model which satisfies the criterion of strong coupling, i.e. with discontinuous modulation function, but with small enough modulation amplitudes, remains necessary to find.

Third, it is known that impurities injected in CDW compounds, at least for the quasi-one dimensional systems, increase the observed electrical threshold fields (see for example \cite{review1D_1_3}). In this respect, the most robust result is the invariance with the doping of the product $\epsilon \times E_c$. We note that this invariance is explicitly satisfied in our scenario since the 'true' threshold field is the local one ($E_{loc.} \sim \epsilon E_c$). However, we do not have at this stage a definitive answer on the exact role of impurities, although they likely soften the dielectric constant.

All these points will be examined in further publications. Eventually, we emphasize that our scenario can be verified by inelastic X-rays or Neutrons experiments by measuring the possible changes in the phonon spectrum under electric field.

\acknowledgments
It is a pleasure for me to acknowledge A. Barbara, D. Le Bolloc'h, E. Lorenzo, D. Mayou and P. Monceau for their support and for many scientific discussions.

\end{document}